\documentclass[article,11pt]{wlscirep}
\usepackage[utf8]{inputenc}
\usepackage[T1]{fontenc}
\usepackage{bm}
\usepackage{smartdiagram}
\usepackage{graphicx}
\usepackage{color}
\usepackage{array, multirow}
\usepackage{amsmath}
\usepackage{amsfonts}
\usepackage{physics}
\usepackage{stackengine}

\usepackage{MnSymbol}
\usepackage{caption}
\usepackage{subcaption}
\usepackage{tikz}
\usetikzlibrary{shapes,arrows}
\usepackage{comment}

\DeclareGraphicsExtensions{.eps}
\begin{document}
\title{Demonstration of quantum advantage by a joint detection receiver for optical communications using quantum belief propagation on a trapped-ion device}
%\author{Conor Delaney, Kaushik Seshadreesan, Ian MacCormack, ..., Saikat Guha, Prineha Narang}
\author[1]{Conor Delaney}
\author[2,$\dagger$]{Kaushik P. Seshadreesan}
\author[1,3,6]{Ian MacCormack}
\author[1, 4]{Alexey Galda}
\author[2]{Saikat Guha}
\author[5,*]{Prineha Narang}
\affil[1]{Aliro Technologies, Inc., Boston Massachusetts 02135, USA}
\affil[2]{College of Optical Sciences, The University of Arizona, Tucson, Arizona 85721, USA.}
\affil[3]{Kadanoff Center for Theoretical Physics, University of Chicago, Chicago, Illinois 60637, USA.}
\affil[4]{James Franck Institute, University of Chicago, Chicago, Illinois 60637, USA.}
\affil[5]{School of Engineering and Applied Sciences,
Harvard University, Cambridge, Massachusetts 02138, USA.}
\affil[6]{Department of Physics, Princeton University, Princeton, New Jersey 08544, USA.}
\affil[$\dagger$]{Email: kaushiksesh@email.arizona.edu}
\affil[*]{Email: prineha@seas.harvard.edu}
\date{\today}

\begin{abstract}
Demonstrations of quantum advantage have largely focused on computational speedups and on quantum simulation of many-body physics, limited by fidelity and capability of current devices. Discriminating laser-pulse-modulated classical-communication codewords at the minimum allowable probability of error using universal-quantum processing presents a promising parallel direction, one that is of both fundamental importance in quantum state discrimination, as well as of technological relevance in deep-space laser communications. Here we present an experimental realization of a quantum joint detection receiver for binary phase shift keying modulated codewords of a 3-bit linear tree code using a recently-proposed quantum algorithm: belief propagation with quantum messages. The receiver, translated to a quantum circuit, was experimentally implemented on a trapped-ion device---the recently released Honeywell LT-1.0 system using ${}^{171}Yb+ $ ions, which possesses all-to-all connectivity and mid-circuit measurement capabilities that are essential to this demonstration. We conclusively realize a previously postulated but hitherto not-demonstrated joint quantum detection scheme, and provide an experimental framework that surpasses the quantum limit on the minimum average decoding error probability associated with pulse-by-pulse detection in the low mean photon number limit. The full joint-detection scheme bridges across photonic and trapped-ion based quantum information science, mapping the photonic coherent states of the modulation alphabet onto inner product-preserving states of single-ion qubits. Looking ahead, our work opens new avenues in hybrid realizations of quantum-enhanced receivers with applications in astronomy and emerging space-based platforms.
\end{abstract}

\maketitle

%\section*{Introduction}
\noindent Optical laser communication is a critical component of future space-based data communications~\cite{deep-space}. It offers significantly higher communication rates compared to traditional radio-frequency systems with lower size, weight and transmission power requirements. \cite{Andrews2008-cx} 
%In order to tap the full potential of laser communications in long-range links such as deep-space communications\cite{Andrews2008-cx}, it is essential to consider the underlying physics. 
An ideal laser pulse is quantum mechanically described by a {\em coherent state} $|\alpha\rangle$ of a spatio-temporal-polarization mode of the quantized electromagnetic field, where $|\alpha|^2$ is the mean photon number~\cite{gerry_knight_2004}. Any two coherent states $|\alpha\rangle,|\beta\rangle$ of a mode are known to be non-orthogonal, i.e., their inner product $\sigma \equiv \langle\alpha|\beta\rangle=\exp(-(|\alpha|^2+|\beta|^2-2\alpha\beta^*)/2)\neq 0$, which fundamentally precludes error-free discrimination of the states~\cite{Helstrom1969-hr}. The minimum achievable probability of error of distinguishing the above two states (assuming they are equally likely to occur) by a physically-realizable receiver as imposed by the laws of quantum mechanics, the so-called {\em Helstrom limit}, is $P_{e,{\rm min}} = \frac{1}{2}[1-\sqrt{1-|\sigma|^2}] = \frac{1}{2}[1-\sqrt{1-e^{-|\alpha - \beta|^2}}]$. This minimum probability of error is in principle attainable {\em exactly} by an all-photonic receiver proposed by Dolinar~\cite{Dolinar1976-og}, which employs a coherent-state local oscillator (LO), a beam splitter, a shot-noise-limited photon detector, and an electro-optic feedback from the detector output to drive an electro-optic modulator (EOM) that controls the amplitude and phase of the LO. Each of these components are readily realizable in a modern optics laboratory. In the context of discriminating more than two coherent states, e.g., discriminating $\left\{|-\alpha\rangle, |0\rangle, |\alpha\rangle\right\}$, or discriminating more than two coherent-state {\em codewords}, e.g., $\left\{|-\alpha\rangle |\alpha\rangle |-\alpha\rangle |\alpha\rangle, |\alpha\rangle |\alpha\rangle |\alpha\rangle |-\alpha\rangle, |\alpha\rangle |-\alpha\rangle |\alpha\rangle |-\alpha\rangle\right\}$, a Dolinar-like all-photonic receiver is not known to achieve the Helstrom limit. There has been a large body of recent work on feedback-based optical receivers for $M$-ary coherent state discrimination~\cite{Chen2012-pm,Becerra2013-ps,Becerra2015-lj,Nair-2014-SWNreceiver}---building upon the ``conditional nulling'' receiver designs for $M$-ary coherent state pulse-position-modulation (PPM)~\cite{dolinar-1983-cpn-PPM, guha-2011-cpn-PPM-theory} and phase-shift-keying (PSK)~\cite{bondurant-93-PSK} alphabets---whose achievable error probability performance were shown to approach the Helstrom limit in the limit of {\em high} mean photon number of the candidate states. In long-range transmissions, photon loss renders the laser communication signals increasingly weak with distance. This attenuation causes the coherent states of the modulation alphabet to become highly non-orthogonal, making it near-impossible to perfectly demodulate coherent state codewords, thereby posing a critical challenge to communicating reliably at a good rate with the distance of the channel. In this regime of low mean photon number per mode at the receiver, a receiver that employs a quantum joint detection measurement on the entire codeword can attain the Helstrom limit of minimum error discrimination of codewords, and there emerges a large gap in the reliable communication rate achievable with a quantum joint-detection receiver (one that would require a true universal-capable quantum processor/computer) versus that achievable with conventional receivers, such as single photon detectors (even if such a conventional receiver is assumed to be operating at its quantum-mechanics-mandated noise limit). Furthermore, no known all-photonic receiver design can attain the multi-hypothesis coherent-state Helstrom limit in this aforesaid regime. However, it is known that if each individual coherent state of the modulation alphabet comprising the received codeword is {\em transduced} into a qubit register one-by-one while maintaining their relative inner product $\sigma$ (hence their quantum-mandated distinguishability), followed by quantum computing on that multi-qubit register, one can achieve the Helstrom limit of telling apart any set of $M \ge 2$ coherent state codewords, exactly~\cite{Da_Silva2013-tk}. 
\newline

\noindent Fundamental limits on the rate of reliable classical communication over a quantum channel with a modulation alphabet consisting of highly non-orthogonal quantum states is given by the Holevo-Schumacher-Westmoreland (HSW) theorem~\cite{Holevo1998-mb, Schumacher1997-cn}, often termed the ``Holevo capacity'', $C$, measured in bits per channel use. For an optical channel with photon loss and thermal noise, each ``use'' of which can be considered to be the transmission of a single spatio-temporal-polarization mode of light under a mean photon number constraint at the transmitter, a coherent state modulation is known to attain the Holevo capacity~\cite{Giovannetti2004-ki,Giovannetti-2014-holevocapacity-thermal-loss}. For any given coherent state modulation alphabet, the structure of the optical receiver governs the achievable reliable communication rate, given by the Shannon capacity associated with a particular receiver. Even though the receiver's job is to tell apart a set of $M=2^{nR}$ product {\em codewords} each being a product state of $n$ coherent states, there is a fundamental gap between the decoding performance achievable with a receiver that detects each received modulated coherent state in the codeword one at a time, versus a receiver that collectively detects the entire codeword using a quantum-enabled processor, thus representing a provable {\it quantum advantage} scenario. A specific realization of such a joint-detection receiver would involve an optical domain quantum pre-processing of the modulated codeword prior to  detection~\cite{Chung2016-to,Dutton2011-zt,Guha2011-nr,Chen2012-pm}. This gap can be quantified in terms of the communication capacity and the average decoding error probability associated with the two types of receivers, and has been shown theoretically\cite{Chung2016-to, Takeoka2014-nb, Giovannetti2004-ki, Guha2011-om,Wilde2013-wz} and verified experimentally\cite{Chen2012-pm, Becerra2013-ps,saikat-nature-photonics}. With a receiver that attains the Holevo capacity, the average probability of error in discriminating the $M=2^{nR}$ codewords can be made to approach zero, as $n$ increases, as long the rate of the code $R < C$.  
\newline

\section*{Fidelity-limited Joint Detection Schemes}

\noindent Recently, a structured design of a quantum joint detection receiver based on an algorithm known as belief propagation with quantum messages (BPQM)\cite{Renes_2017} was proposed to discriminate binary PSK (BPSK)-modulated coherent-state codewords of an exemplary $5$-bit linear tree code. It was shown not only to surpass the performance of the best-possible conventional receiver that detects the received coherent state pulses one at a time, but to attain the quantum limit on the minimum average decoding error probability\cite{rengaswamy2020quantummessagepassing, RSGP20}, the codeword Helstrom limit. The design of the receiver readily translates into a low-depth quantum circuit realizable on current quantum devices, which are designed for complex algorithms\cite{11-qubitcomputer,chem_review,trappionqc,Monz1068,zoller-open-system-sim,DJalgo}. We specifically realize sections of a joint detection receiver circuitry on Honeywell's LT-1.0 trapped-ion processor, leveraging all-to-all gate connectivity and mid-circuit measurements. The necessity of these mid-circuit measurements, currently not viable on superconducting quantum devices, makes trapped-ion processors the ideal platform for this demonstration. We also propose a concrete transduction mechanism to couple the states $\left\{|\alpha\rangle, |-\alpha\rangle\right\}$ of the BPSK alphabet to (one of two states of) a single trapped-ion qubit. Although the coupling is not physically realized, when coupling inefficiencies are accounted for in the realization of the joint detection receiver circuitry, it still demonstrates a fundamentally improved performance in the decoding error probability achievable over {\em any} receiver that demodulates the BPSK pulses in the codeword blocks one at a time. This includes all conventional optical receivers such as homodyne detection, heterodyne detection, and direct detection receivers (for example, superconducting nanowire single photon detectors), as well as the Dolinar receiver~\cite{Dolinar1976-og}.\\

\begin{figure}
    \centering
    \includegraphics[width=\textwidth]{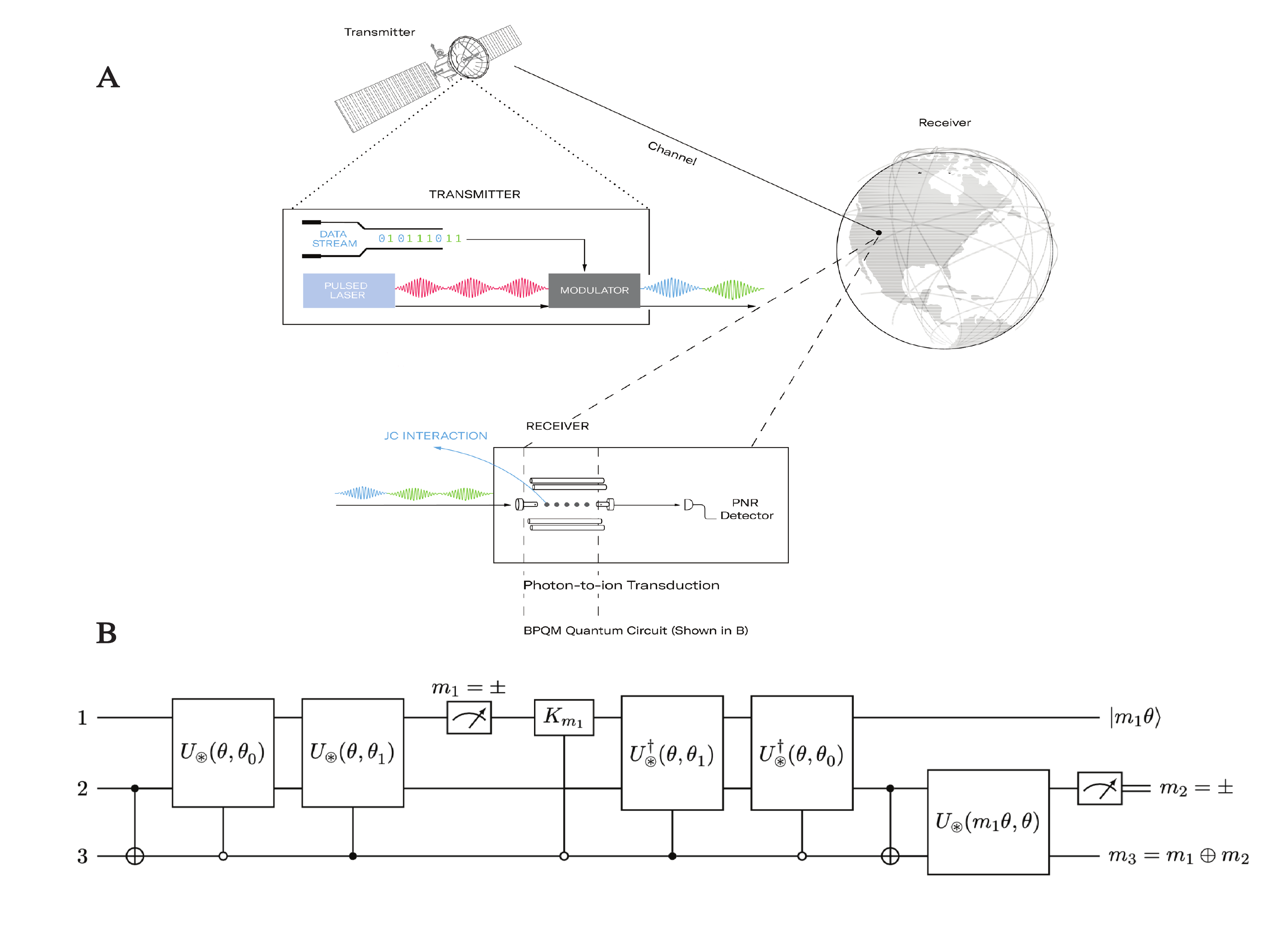}
    %\begin{subfigure}[b]{.45\textwidth}
    %    \label{fig:schematic}
    %    %\caption{ }
    %    %\includegraphics[width=\textwidth]{schematic_mini.pdf}
    %    \tikzstyle{block} = [rectangle, draw, fill=gray!20, 
    %        text width=4em, text centered, rounded corners, minimum height=4em]
    %    \tikzstyle{blocktwo} = [rectangle, draw, fill=yellow!50, 
    %        text width=4em, text centered, rounded corners, minimum height=4em]
    %    \tikzstyle{line} = [draw, -latex']
    %    \begin{tikzpicture}[node distance = 3cm]
    %    \node [block] (signal) {Signal};
    %    \node [block, right of=signal] (receive) {Receiver};
    %    \node [blocktwo, right of=receive] (hardware) {Trapped Ion Hardware};
    %    \path [line] (signal) -- (receive);
    %    \path [line] (receive) -- (hardware);
    %    \end{tikzpicture}
    %\end{subfigure}
    %\hspace{10}
    %\begin{subfigure}[b]{.4\textwidth}
    %    %\caption{ }
    %    \begin{center}
    %    \begin{tikzpicture}
    %   \node[draw,circle,fill=blue!30] (x1) at (0,0) {$x_1$};
    %    \node[draw,rectangle,fill=red!50] (c) at (0,-1) {$c$};
    %    \node[align=center] (PC) at (+1.75,-1) {$x_1\oplus x_2\oplus x_3=0$};
    %    \node[draw,circle,fill=blue!30] (x2) at (-0.75,-2) {$x_2$};
    %    \node[draw,circle,fill=blue!30] (x3) at (+0.75,-2) {$x_3$};
    %    \draw (x1) -- (c);
    %    \draw (x2) -- (c);
    %    \draw (x3) -- (c);
    %    \end{tikzpicture}
    %    \label{fig:factorgraph}
    %    \end{center}
    %\end{subfigure}
    %\vspace{20}
    %\vfill
    %\begin{subfigure}[b]{\textwidth}
    %    \includegraphics[width=.95\textwidth]{3bitcode_full_circuit.pdf}
    %    %\caption{ }
    %    \label{fig:circ_diagram}
    %\end{subfigure}
    \caption{The schematic and operation of the quantum joint-detection receiver for decoding a 3-bit laser-modulated 
    %linear tree 
    code. The encoded photonic information is efficiently decoded using a trapped ion quantum computer (A), which executes the 3-qubit BPQM algorithm circuit (B). Notation and circuit structure discussed in the Methods section.
    %This circuit requires mid-circuit measurement capabilities which will decode the incoming message.
    }
    \label{fig:full_schematic}
\end{figure}

\noindent Realization of a true joint-detection receiver in the near-term requires heterogeneous quantum hardware, namely trapped-ion and photonic systems, in close coupling with theoretical efforts to map across them \cite{ULTRAFASTPhysRevLett.112.250501,long_distance_quantum_comm}. The ability to perform the BPQM decoding algorithm, which effects a joint measurement to distinguish the photonically-encoded messages, is a single step in the overall scheme. The general overview of the scheme is presented in Fig. \ref{fig:full_schematic}A, which shows a long-distance photonic communication being received and decoded. The receiver here requires a method for transduction from the photonic information domain into the trapped-ion quantum device, as well as quantum hardware with minimal noise to run the decoding efficiently and reliably. In this work we focus on the use of trapped-ion devices, specifically the Honeywell LT-1.0 system, although in theory this could be realized with any quantum computer with low enough noise and the ability to perform mid-circuit measurements. The full joint-detection scheme relies on leveraging both photonic and trapped-ion based information; each of which has been explored in depth \cite{tele-review,ion-photon04,modular_entanglement,PhysRevA.89.022317} and will be addressed next.

\subsection*{BPQM Decoding}
To decode laser communication messages with BPQM, we first present the  specific implementation of the algorithm. The photonic input states, namely BPSK coherent states $|\pm\beta\rangle,$ are represented as qubit states $|\pm\theta\rangle,$ by the mapping
\begin{gather}
|\pm\beta\rangle\rightarrow |\pm\theta\rangle\equiv\cos\left(\frac{\theta}{2}\right)|0\rangle\pm\sin\left(\frac{\theta}{2}\right)|1\rangle,
\end{gather}
such that $\sigma=\langle+\beta|-\beta\rangle=\langle+\theta|-\theta\rangle=\cos\theta\neq 0.$ The task is to find an efficient decoding algorithm that can discriminate codewords constructed using the alphabet defined by these non-orthogonal quantum states.
The decoder based on the BPQM algorithm~\cite{Renes_2017} was recently analyzed by Rengaswamy et al \cite{rengaswamy2020quantummessagepassing} for a 5-bit linear tree code, where in noiseless simulations it was shown to surpass the classical bound for decoding error rates at low mean photon numbers. This was followed by a quantum gate decomposition for the various unitary operators described, which provides a starting point for implementation on a real device. These quantum gates effectively perform belief-propagation by combining the beliefs at the nodes of the factor graph of the code before iteratively passing on the updated beliefs until the message is jointly decoded, just as in the classical belief-propagation algorithm. The difference here is the leveraging of the quantum regime, where the decoder passes quantum ``beliefs'' and jointly processes the quantum information present in the symbols before measuring them individually. This allows us to bypass the inevitable loss of information that comes from measuring the individual symbols first followed by processing the detection outcomes classically. For an example 3-bit code $\mathcal{C}$, we arrive at the circuit for the BPQM-based decoder based on the development in Ref.\cite{rengaswamy2020quantummessagepassing}, shown in Fig.\ref{fig:full_schematic}B. Further description of the code $\mathcal{C}$ and the implementation of the BPQM algorithm for the decoder can be found in the Methods section.

\subsection*{Photon-to-Ion Transduction}

Mapping the binary BPSK coherent state alphabet onto one of two single qubit states --- henceforth called the transduction step --- is necessary to fully realize the joint detection receiver. In this step, it is essential that the inner product between the non-orthogonal binary states of the qubits remain the same as that of the received coherent states (under ideal conditions). For coherent states $|\pm\alpha\rangle$ transmitted over a lossy channel of transmissivity $\eta$, the received states are $|\pm \beta \rangle=|\pm \sqrt{\eta}\alpha \rangle$ with an overlap of
\begin{equation}
    \langle+\beta|-\beta\rangle= e^{-2  |\beta|^2} %=\langle+\sqrt{\eta}\alpha|-\sqrt{\eta}\alpha\rangle 
    = e^{-2 \eta |\alpha|^2}=e^{-2N},
\end{equation}
$N$ being the received mean photon number.
Below we outline a process of performing the aforementioned transduction using the simple and experimentally realizable Jaynes-Cummings interaction between a qubit and a single bosonic mode~\cite{PhysRevA.52.4214,RevModPhys.75.281}.
\newline

\noindent Based on prior results from \cite{2018NJPh...20d3005H}, we can start by writing down the product state of a single photon mode and a two level atom (a trapped-ion for our purposes), where the photon mode has been initialized in one of the following two coherent states
\begin{equation}
|\pm \beta \rangle = \sum_n  e^{-|\beta|^2/2} \frac{(\pm \beta)^n}{\sqrt{n!}} |n \rangle,
\end{equation}
the atom is initialized in its ground state $|0 \rangle$, and the two evolve with the following time-dependent Hamiltonian.
\begin{equation}
    H = \hbar \Omega(t) (\sigma_+ a + \sigma_- a^{\dagger}).
\end{equation}
Here $\sigma_\pm$ are the raising and lowering operators for the trapped-ion qubit, and $a$ and $a^\dagger$ are photon creation and annihilation operators. Time evolving the initial product state with the above Hamiltonian we get the following entangled state:
\begin{equation}
    |\Psi^\pm (t) \rangle = \sum_n  \big ( \cos (\Phi \sqrt{n}) \beta^\pm_n |0,n \rangle - i \sin (\Phi \sqrt{n+1}) \beta^\pm_{n+1}|1,n \rangle \big ),
\end{equation}
where
\begin{equation}
    \Phi(t) = \int_0^t dt' \Omega(t')
\end{equation}
and
\begin{equation}
    \beta^\pm_n = e^{-|\beta|^2/2} \frac{(\pm \beta)^n}{\sqrt{n!}}.
\end{equation}
Since this time evolution is unitary, one can verify that the state remains normalized. We now perform a projective measurement on the photon in order to obtain the desired qubit state. The inner product of the two binary qubit states after will depend on the photon measurement result. Since $n=0$ is the most likely measurement outcome, we will ultimately tailor the interaction $\Omega$ accordingly, so that an $n=0$ measurement heralds a successful transduction. The probability of achieving an $n=0$ measurement result can be expressed as
\begin{equation}
    P(n=0) = e^{-|\beta|^2} (1+\sin^2 \Phi(t) \beta^2),
\end{equation}
and the resulting normalized state of the qubit will be 
\begin{equation}
    P|_{n=0} |\Psi^\pm (t) \rangle = \frac{1}{\sqrt{1+\sin^2 \Phi(t) \beta^2}} \left ( |0 \rangle \mp i \sin \Phi(t) \beta |1 \rangle \right)
\end{equation}
For a given $\beta$, if we were to pick $\Phi$ so that the inner products of the optical BPSK states match those of the post-transduction states of the trapped ion qubit, we would need to satisfy
\begin{equation}
  ( \langle \Psi^- (t) | P|_{n=0})  P|_{n=0} |\Psi^+ (t) \rangle= \langle -\beta | \beta \rangle = e^{-2|\beta|^2},
\end{equation}
which would imply the following must hold:
\begin{equation}
    \sin \Phi = \frac{1}{\beta} \sqrt{\tanh |\beta|^2}.\label{inner-pdt_preserving_Phi}
\end{equation}
Thus, we can tailor the time-dependent interaction $\Omega(t)$ so that its integral $\Phi$ satisfies the above relation \cite{PhysRevA.71.023811}. Plugging this condition into the $n=0$ measurement probability, we can compute the probability of a successful transduction (not accounting for noise) to be as follows:
\begin{equation}
    P(n=0) = e^{-|\beta|^2} (1+\tanh |\beta|^2)=e^{-\eta|\alpha|^2} (1+\tanh (\eta|\alpha|^2)).\label{p_her_inner-pdt_preserving_Phi}
\end{equation}
Note that when $\eta \ll 1$, the above probability decreases very slowly with the transmitted coherent amplitude $\alpha$ since measuring $n=0$ will be highly probable.
\newline

\begin{figure}
    \centering
    \includegraphics[width=.55\textwidth]{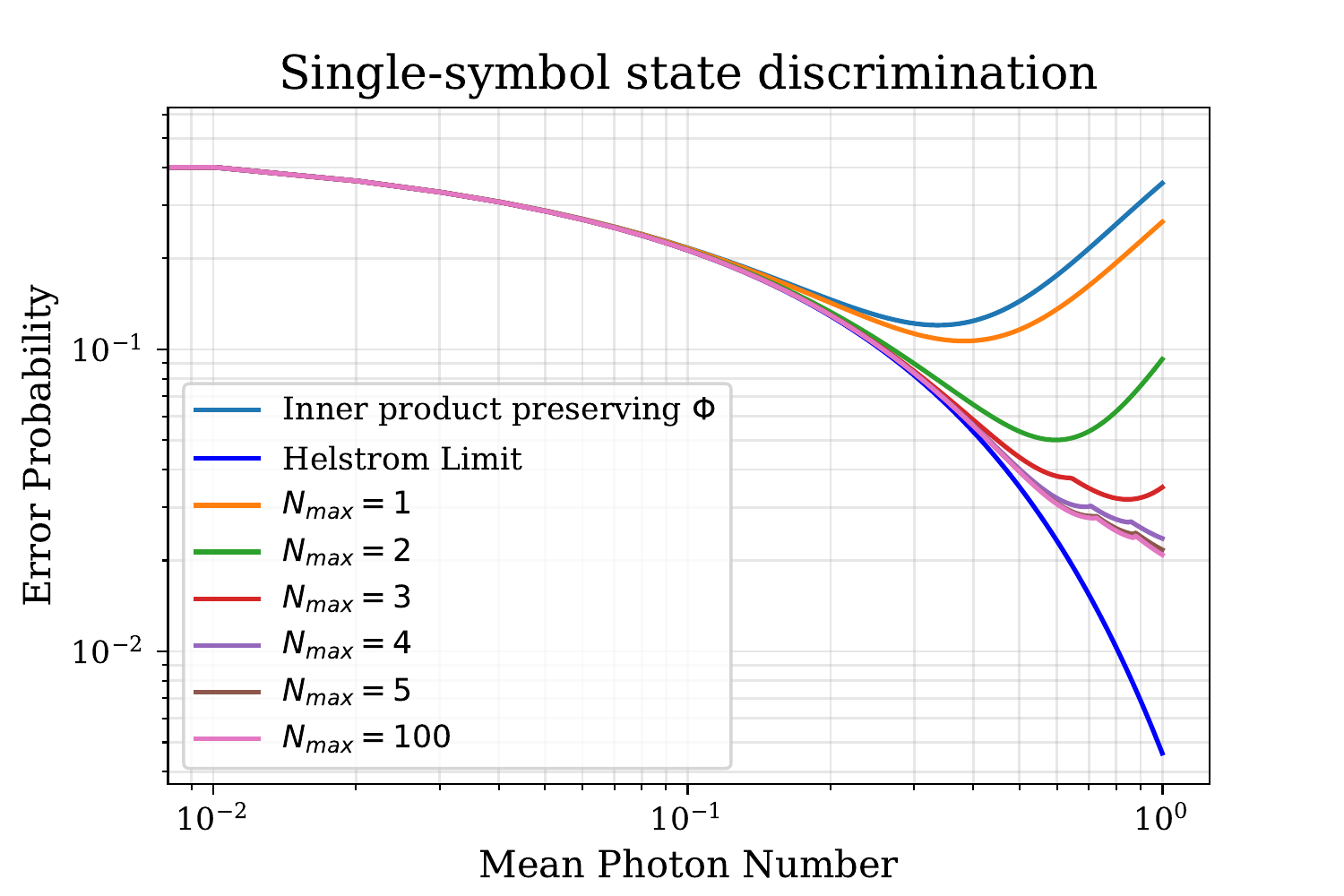}
    \caption{The single-symbol state discrimination error probability as a function of the received mean photon number. Helstrom bound (dark blue) shows the absolute minimum error probability of discriminating the BPSK alphabet binary coherent states in the optical domain, whereas inner product preserving $\Phi$ (light blue) and optimized $\Phi$ for various points of truncation in the sum \ref{prob_sum} show the overall average error probability of optical BPSK state discrimination, using our receiver. The photon-to-ion transduction step uses two different choices for $\Phi$, e.g., $\Phi$ given by Eq.~\ref{inner-pdt_preserving_Phi} for the inner product preserving transduction and $\Phi=\pi/2$ for transduction that results in the optimal overall average error probability when we truncate \ref{prob_sum} after $n=1$. All of the above assume ideal quantum logic gates and measurements once in the ion domain.}
    \label{fig:error_prob_chart}
\end{figure}

\noindent With the inclusion of the transduction step, the overall average probability of successful discrimination of the BPSK alphabet binary coherent states is given by the probability that the $n=0$ outcome occurs in the transduction step (heralding probability of successful transduction), multiplied by the maximum success probability of discriminating the two non-orthogonal qubit states within the trapped ion quantum computer given by $1-P_{e,\textrm{min}}$, where $P_{e,\textrm{min}}$ is the Helstrom limit associated with error probability of discriminating the transduced qubit states (here we assume that the quantum gates and measurements on that trapped-ion qubit are perfect). Thus, the overall average error probability is most generally:
\begin{equation}
    P_{\rm{error}}= 1-P(n=0)(1- P_{e, {\rm min}} )=1- \frac{e^{-|\beta|^2} }{2} (1+\sin^2 \Phi(t) \beta^2) \left[1+ \sqrt{1-\frac{(1-\sin^2 \Phi(t) \beta^2)^2}{(1+\sin^2 \Phi(t) \beta^2)^2} }\right].\label{transduced P_error}
\end{equation}
For a $\Phi$ chosen according to Eq.~\ref{inner-pdt_preserving_Phi}, the $P_{\rm{error}}$ of Eq.~\ref{transduced P_error} corresponds to the discrimination error probability associated with an inner-product preserving transduction step.
\newline
\newline
\noindent If we relax the requirement to preserve the inner product before and after transduction, we can obtain even better performance of overall discrimination of the BPSK coherent state alphabet states. By controlling the interaction time, and hence $\Phi$, we can make the inner product of the transduced states smaller than that of the optical BPSK states, which increases the heralded success probability of state discrimination in the ion domain. But this comes at the cost of a smaller heralding probability $P(n=0)$, which ensures that the product, i.e., the overall average error probability, remains below the Helstrom limit associated with discriminating the original BPSK binary coherent states. We can minimize $P_{\rm{error}}$ of Eq.~\ref{transduced P_error} with respect to $\Phi$ to find the minimum overall probability of error. The optimal choice of $\Phi$, interestingly, works out to be not dependent on $\beta$, as shown below. The minimum occurs when $\Phi(t)=\frac{\pi}{2}$ and is given by
\begin{equation}
    P_{\rm{error}}\geq 1- \frac{e^{-|\beta|^2} }{2} (1+|\beta|)^2
\end{equation}
For $\Phi= \frac{\pi}{2}$, the inner product of the qubit embeddings of the coherent states is 
\begin{equation}
    \langle \Psi^- |\Psi^+ \rangle = \frac{1-|\beta|^2}{1+|\beta|^2}
\end{equation}
which, one can verify, is always smaller than $e^{-2|\beta|^2}$. Despite this, we have actually increased the average probability of successfully discriminating the coherent-state BPSK alphabet, by optimally choosing $\Phi$. The improvement when compared to the $\Phi$ of  Eq.~\ref{inner-pdt_preserving_Phi} corresponding to inner-product preserving transduction is shown in Fig. \ref{fig:error_prob_chart}. The Helstrom limit associated with discriminating the original BPSK alphabet binary coherent states is also plotted for comparison. In the following sections we will consider the exact inner product scenario (Eq.~\ref{p_her_inner-pdt_preserving_Phi}) as our probability of successful transduction, but the above discussion shows that the experimentally-obtained performance reported in this paper can only improve further if the optimal $\Phi$ is chosen for the transduction step.
\newline

\noindent If we have photon number resolving (PNR) detection available, we can resolve higher (non-zero) values of $n$, and the average error probability of discriminating the BPSK coherent states attained by our transduction method followed by an ideal trapped-ion quantum processor is given by:
\begin{equation}
    P_{e,{\rm receiver}}(\Phi) = 1- \frac{1}{2} \sum_n P_n(\Phi) \lbrack 1+\sqrt{1-\sigma_n^2} \rbrack,
    \label{prob_sum}
\end{equation}
which, as before, can be minimized by optimally choosing $\Phi$. Above, $\sigma_n$ is the inner product between the two possible ion states heralded by a measurement of $n$ photons, which is

\begin{equation}
    \sigma_n= \frac{\cos^2(\sqrt{n}\Phi) - \frac{\beta^2}{n+1} \sin^2(\sqrt{n+1}\Phi) }{\cos^2(\sqrt{n}\Phi) + \frac{\beta^2}{n+1} \sin^2(\sqrt{n+1}\Phi)}.
\end{equation}

Though the terms in the sum are rather complicated, for a given value of $\beta$, one can easily numerically minimize the above function, and include arbitrarily many terms. This has been done for several different levels of series truncation in Fig. \ref{fig:error_prob_chart}. This would ensure optimized performance assuming the availability of PNR detection. 
\newline

%\noindent As an alternative to the transduction scheme presented above based on the Jaynes-Cummings interaction, one could exploit photonic interconnect technology developed as a means of building a large, modular trapped-ion computer \cite{ion-photon04,modular_entanglement,PhysRevA.89.022317}. See, for example, Ref. \cite{2012arXiv1208.0391M} for details of how such an interconnect would work. This, however, would be a significantly more difficult approach than implementation of the Jaynes-Cummings Hamiltonian due to the necessity of single rail logic and a much lower $P(n=0)$ value.

\section*{BPQM on the Honeywell LT-1.0 Trapped-Ion Processor}
Next we present the demonstration of the BPQM algorithm on a recently developed quantum device. The implementation on a currently available QPU provides a performance standard and outlook for these joint-detection receivers based on the scheme set forth in this work. For this experiment, we utilized the Honeywell LT-1.0 trapped-ion device, which uses ${}^{171}Yb+ $ ions. The choice of the device was motivated by the unique combination of high-fidelity quantum gates, all-to-all qubit connectivity afforded by trapped-ion architecture, and the unique capability to perform mid-circuit measurements on selected qubits to condition subsequent gate operations on their measurement outcomes. The all-to-all connectivity enables a number of circuit optimizations that allow the avoidance of costly SWAP gates, resulting in the compact decomposition of the circuit depicted in Fig. \ref{fig:full_schematic}B, which requires 81 two-qubit M{\o}lmer-S{\o}renson-like ZZ gates\cite{PhysRevLett.82.1835}. In the absence of a physical implementation of the photon-to-ion transduction, the initial states of the qubits are prepared directly based on the chosen codeword for every given run rather than created as a result of the photon projective measurement. With the exception of the noisy simulation, the data points were taken assuming lossless transduction. While the proposed transduction scheme has not been exactly experimentally implemented, the Jaynes-Cummings coupling already serves as a reasonable model of the laser-ion interaction in current trapped-ion devices \cite{RevModPhys.85.1083}, making it a highly possible near-term development.
\newline

%state-of-the-art in photonic interconnect for modular ion traps have recently shown fidelities up to $94\%$ \cite{PhysRevLett.124.110501}, which further gives confidence in the feasibility of technological realizations of such a scheme. 
\begin{figure}
    \centering
    \includegraphics[width=\textwidth]{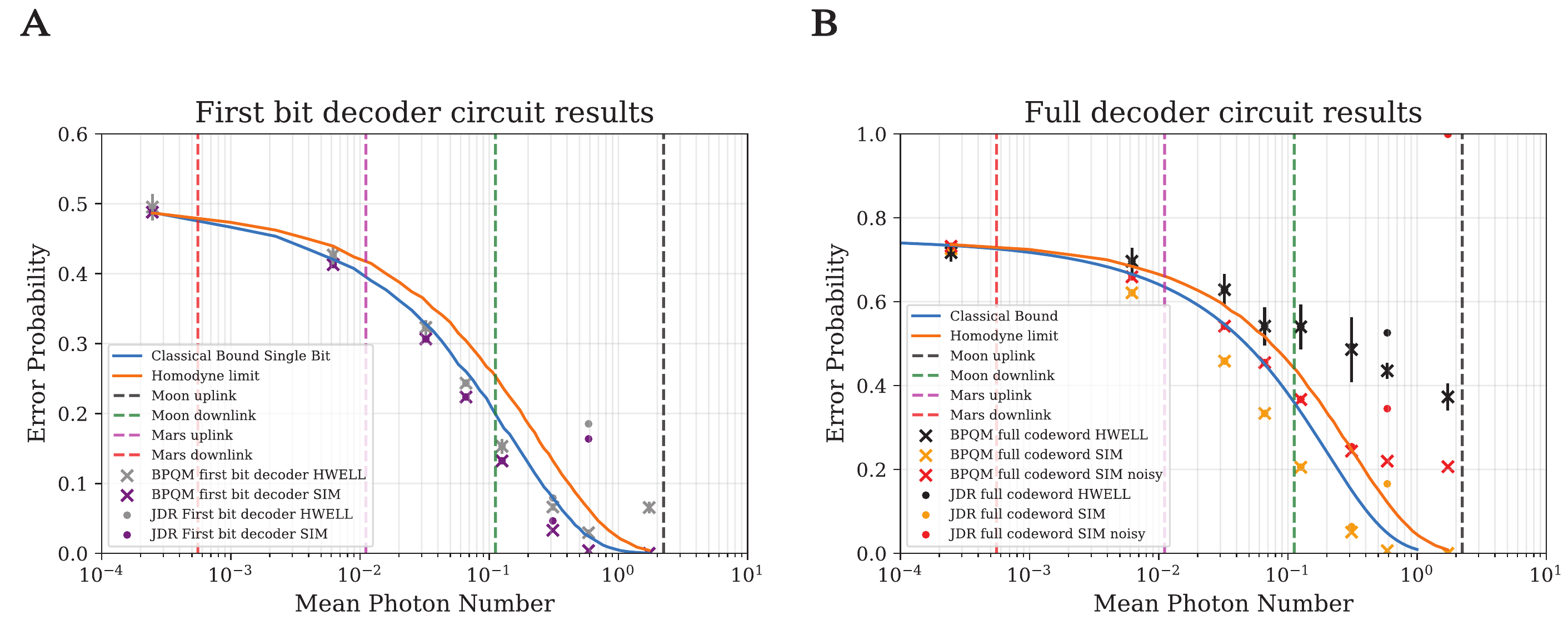}
    \caption{Experimental results for the first-bit (A) and full 3-bit (B) decoder with four codewords, with 1000 and 512 shots per run, respectively. The ``classical" bound represents the error probability associated with measuring the individual symbols in the photonic domain sans the trapped-ion receiver using the quantum-optimal Helstrom measurement followed by classical maximum likelihood decoding (blue line). The Homodyne limit corresponds to a practical classical bound, where the Helstrom measurements are replaced by homodyne measurements (orange line). "BPQM" points represent circuit runs as-is with perfect transduction assumed, whereas "JDR" points account for the probability of successful transduction based on our scheme (the cube of Eq.~\ref{p_her_inner-pdt_preserving_Phi} in (B), to account for three qubits). Experimental error probabilities of decoding with the trapped-ion receiver  (grey crosses first bit, black crosses full decoder) are averaged over four codewords, with error bars for standard deviation. Noiseless (purple crosses for the first bit, yellow crosses for the full decoder) and noisy (red crosses) simulation results are shown for comparison.}
    \label{fig:decoder_results}
\end{figure}
%\begin{figure}
%    \centering
%    \includegraphics[width=.7\textwidth]{single_decoder.pdf}
%    \caption{Results for the single bit decoder. Points are plotted as the average of the %performance of each of the four codewords; all Honeywell circuits run with 1000 shots. Error bars %represent the standard deviation of the performance of the four codewords. Blue classical bound %represents the ideal classical limit (helstrom) whereas the Homodyne limit represents the practical %classical bound.}
%    \label{fig:single_bit_decoder}
%\end{figure}

\noindent As a first step, we look at decoding only the first bit of the full codeword. For this we are able to use an abbreviated version of the circuit that is truncated at the first measurement on the top qubit. This gives us an estimate of how the decoder and the $U_{\ostar}$ unitary gates are performing on the device without immediately evaluating the longer gate depth of the full decoder. This significantly reduces the gate count, allowing us to exceed the classical bound for a range of low received mean photon numbers, shown in Fig. \ref{fig:decoder_results}A. While this demonstration shows relatively modest improvements when compared to classical approaches, these points give us confidence in the implementation of the $U_{\ostar}$ blocks and allow us to move forward to the full circuit.
\newline

\begin{figure}
    \centering
    \includegraphics[width=.75\textwidth]{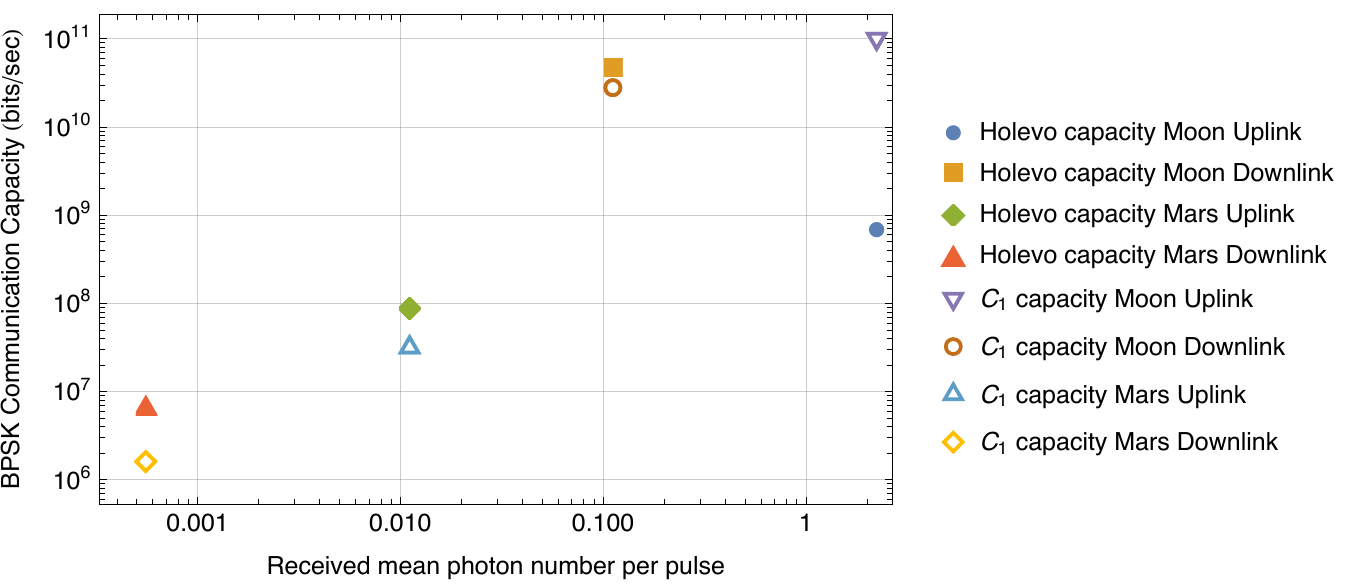}
    \caption{Link budgets based on LLC specs for an uplink and downlink. The Holevo capacities plotted here are ``achievable'' Holevo capacity that denote the quantum bound for classical communication capacity using joint detection of symbols via probabilistic photon-to-ion transduction. The $C_1$ capacities are the capacities associated with symbol-by-symbol optical detection for the BPSK scheme at the given link distances.}
    \label{fig:link_plot}
\end{figure}

\noindent In the full decoder circuit, the codeword output is determined by both the intermediate and final measurements. This circuit gives us an accurate look at the performance of BPQM on current devices. Noiseless simulations are shown to beat the classical bound for error probability of decoding over a range of low mean photon numbers, for values within the moon to mars down-link regime in Fig. \ref{fig:decoder_results}B. On the other hand, running the circuit on hardware produces a curve that trends at  and peeks below the classical bound at very low mean photon numbers, presenting a quantum advantage. We note that the hardware curve shows an anomalous behavior with increasing mean photon number, where it trends worse than the classical bound. This is due to the infidelities associated with initializing the trapped-ion qubits in states $|\pm \theta\rangle$ corresponding to large coherent amplitudes $\beta$ in the states $|\pm\beta\rangle$. Additionally, the ``JDR'' points diverge from the ``BPQM'' points as the mean photon number becomes larger due to a rapidly decreasing probability of successful transduction. However, when we consider the advantage scenario of low mean photon numbers, our noisy simulations to find the operating fidelities of one and two qubit depolarizing error that can bring us to the classical bound show a  `checkpoint' that can drive future experimental pushes. This is shown in Fig. \ref{fig:decoder_results}B along with the experimental data. We consider a simple depolarizing error model with 1 \& 2 qubit depolarizing noise set to $0.0001$ and $0.005$, respectively, while the photonic input state preparation was given a $0.0001\%$ fail rate based on values of JC error rates within existing ion traps being approximately equivalent to a single qubit gate. This checkpoint is intended to be viewed as a more general noise level regime rather than a specific benchmark, but it shows a clear path toward achieving fully useful quantum advantage in the low photon number regime. It is thus clear that the BPQM algorithm is mainly limited by gate fidelity in achieving a lower error probability for decoding messages for the types of channels highlighted. 
\newline

\noindent Of particular interest in Fig. \ref{fig:decoder_results} is the projected mean photon number corresponding to lunar and a future Mars link based on the specifications of optical elements used in NASA's 2013 lunar laser communication demonstration\cite{10.1117/12.2045508} (LLCD) experiment. For example, the Mars uplink corresponds to a received mean photon number per pulse of $10^{-2}$. Notably, at this mean photon number, the BPQM-based receiver ideally surpasses the classical limit in the average error probability of decoding the codes of the example 3-bit code by approximately 2-3\%. At the same mean photon number, by choosing a different code --- one that achieves the Holevo capacity--- it would be possible to reliably communicate at rates $~5\times$ the best possible rate for communication with classical decoders, as depicted in Fig.~\ref{fig:link_plot}. Note that the capacities plotted in Fig.~\ref{fig:link_plot} are ``achievable'' Holevo capacities that take into account the probability of successful photon-to-ion transduction, which still allows us better rates for all links except the moon uplink due to its higher mean photon number and thus low $P(n=0)$.

\section*{Conclusions and Outlook}

Here, we conclusively realize a previously postulated joint quantum detection scheme on a trapped-ion quantum device, and show an experimental framework to surpass the quantum limit on the minimum average decoding error probability in the low-photon limit. By leveraging a combination of mid-circuit measurement-enabled experiments, the connectivity of trapped-ion devices, and a mapping of the relevant photonic coherent states onto inner product-preserving single qubit states, our work shows a quantum joint detection receiver for a 3-bit BPSK modulated linear tree code using the BPQM algorithm. Continued reduction of trapped-ion device error rates --- particularly two-qubit gate infidelities and mid-circuit measurement-induced crosstalk error rates --- will push the noise boundary such that future experiments of this scheme can reliably exceed the classical bound for low photon numbers. Further, as gate fidelities improve, the post-measurement error mitigation techniques and gate decomposition optimizations presented here will give us a distinct path towards exceeding the classical bound for decoding in general joint detection schemes with a larger codebook.
\newline

\noindent The joint-detection receiver protocol shown here provides an additional impetus for the development of photonic transduction in trapped-ion hardware in the near-future. Photonic interconnects are already under development for the purpose of constructing modular trapped-ion architectures \cite{2012arXiv1208.0391M,coupling_patent}, and the basic functionality can in principle be extended to connect a trapped-ion device to a photonic quantum device. Regardless, photon-to-ion transduction will be an essential feature of any technological realization of BPQM. Finally, we highlight the promise of such schemes for deep-space communications and upcoming space missions, a dedicated Mars link, and for advances in astronomy.

\section*{Acknowledgements}
This work is supported by Air Force STTR grant numbers FA8750-20-P-1721 and FA8750-20-P-1704. KPS and SG
acknowledge support of a National Science Foundation (NSF) project ``CIF: Medium: Iterative Quantum LDPC
Decoders”, award number: 1855879, and the Office of Naval Research (ONR) MURI project on ``Optical Computing",
grant number N00014-14-1-0505. The authors gratefully acknowledge the entire Honeywell Quantum Solutions team, especially Dr. Brian Neyenhuis, for helpful discussions and support in running these experiments. The authors thank Dr. Michael Fanto (AFRL) as well as Steve Willis and Dr. Nidhi Aggarwal (Aliro Technologies) for helpful feedback on the work and manuscript. KPS thanks Dr. Narayanan Rengaswamy for helpful discussions.

\bibliography{main.bib}

\begin{thebibliography}{10}
\urlstyle{rm}
\expandafter\ifx\csname url\endcsname\relax
  \def\url#1{\texttt{#1}}\fi
\expandafter\ifx\csname urlprefix\endcsname\relax\def\urlprefix{URL }\fi
\expandafter\ifx\csname doiprefix\endcsname\relax\def\doiprefix{DOI: }\fi
\providecommand{\bibinfo}[2]{#2}
\providecommand{\eprint}[2][]{\url{#2}}

\bibitem{deep-space}
\bibinfo{author}{Deutsch, L.~J.}
\newblock \bibinfo{journal}{\bibinfo{title}{Towards deep space optical
  communications}}.
\newblock {\emph{\JournalTitle{Nature Astronomy}}}
  \textbf{\bibinfo{volume}{4}}, \bibinfo{pages}{907--907},
  \doiprefix\url{10.1038/s41550-020-1193-1} (\bibinfo{year}{2020}).

\bibitem{Andrews2008-cx}
\bibinfo{author}{Andrews, K.}, \bibinfo{author}{Divsalar, D.},
  \bibinfo{author}{Dolinar, S.}, \bibinfo{author}{Hamkins, J.} \&
  \bibinfo{author}{Pollara, F.}
\newblock \bibinfo{journal}{\bibinfo{title}{Design and standardization of
  {Low-Density} {Parity-Check} codes for space applications}}.
\newblock {\emph{\JournalTitle{SpaceOps 2008 Conference}}}
  (\bibinfo{year}{2008}).

\bibitem{gerry_knight_2004}
\bibinfo{author}{Gerry, C.}, \bibinfo{author}{Knight, P.} \&
  \bibinfo{author}{Knight, P.~L.}
\newblock \emph{\bibinfo{title}{Introductory quantum optics}}
  (\bibinfo{publisher}{Cambridge university press}, \bibinfo{year}{2005}).

\bibitem{Helstrom1969-hr}
\bibinfo{author}{Helstrom, C.~W.}
\newblock \bibinfo{journal}{\bibinfo{title}{Quantum detection and estimation
  theory}}.
\newblock {\emph{\JournalTitle{Journal of Statistical Physics}}}
  \textbf{\bibinfo{volume}{1}}, \bibinfo{pages}{231--252}
  (\bibinfo{year}{1969}).

\bibitem{Dolinar1976-og}
\bibinfo{author}{Dolinar, S.~J.}
\newblock \emph{\bibinfo{title}{A class of optical receivers using optical
  feedback}}.
\newblock Ph.D. thesis, \bibinfo{school}{Massachusetts Institute of Technology}
  (\bibinfo{year}{1976}).

\bibitem{Chen2012-pm}
\bibinfo{author}{Chen, J.}, \bibinfo{author}{Habif, J.~L.},
  \bibinfo{author}{Dutton, Z.}, \bibinfo{author}{Lazarus, R.} \&
  \bibinfo{author}{Guha, S.}
\newblock \bibinfo{journal}{\bibinfo{title}{Optical codeword demodulation with
  error rates below the standard quantum limit using a conditional nulling
  receiver}}.
\newblock {\emph{\JournalTitle{Nature Photonics}}}
  \textbf{\bibinfo{volume}{6}}, \bibinfo{pages}{374--379}
  (\bibinfo{year}{2012}).

\bibitem{Becerra2013-ps}
\bibinfo{author}{Becerra, F.~E.} \emph{et~al.}
\newblock \bibinfo{journal}{\bibinfo{title}{Experimental demonstration of a
  receiver beating the standard quantum limit for multiple nonorthogonal state
  discrimination}}.
\newblock {\emph{\JournalTitle{Nature Photonics}}}
  \textbf{\bibinfo{volume}{7}}, \bibinfo{pages}{147--152}
  (\bibinfo{year}{2013}).

\bibitem{Becerra2015-lj}
\bibinfo{author}{Becerra, F.~E.}, \bibinfo{author}{Fan, J.} \&
  \bibinfo{author}{Migdall, A.}
\newblock \bibinfo{journal}{\bibinfo{title}{Photon number resolution enables
  quantum receiver for realistic coherent optical communications}}.
\newblock {\emph{\JournalTitle{Nature Photonics}}}
  \textbf{\bibinfo{volume}{9}}, \bibinfo{pages}{48--53} (\bibinfo{year}{2015}).

\bibitem{Nair-2014-SWNreceiver}
\bibinfo{author}{Nair, R.}, \bibinfo{author}{Guha, S.} \& \bibinfo{author}{Tan,
  S.-H.}
\newblock \bibinfo{journal}{\bibinfo{title}{Realizable receivers for
  discriminating coherent and multicopy quantum states near the quantum
  limit}}.
\newblock {\emph{\JournalTitle{Phys. Rev. A}}} \textbf{\bibinfo{volume}{89}},
  \bibinfo{pages}{032318}, \doiprefix\url{10.1103/PhysRevA.89.032318}
  (\bibinfo{year}{2014}).

\bibitem{dolinar-1983-cpn-PPM}
\bibinfo{author}{Dolinar, S.~J.}
\newblock \bibinfo{journal}{\bibinfo{title}{A near-optimum receiver structure
  for the detection of m-ary optical ppm signals}}.
\newblock {\emph{\JournalTitle{JPL TDA Progress Report}}}
  \bibinfo{pages}{42–--72} (\bibinfo{year}{1983}).

\bibitem{guha-2011-cpn-PPM-theory}
\bibinfo{author}{Guha, S.}, \bibinfo{author}{Habif, J.~L.} \&
  \bibinfo{author}{Takeoka, M.}
\newblock \bibinfo{journal}{\bibinfo{title}{Approaching helstrom limits to
  optical pulse-position demodulation using single photon detection and optical
  feedback}}.
\newblock {\emph{\JournalTitle{Journal of Modern Optics}}}
  \textbf{\bibinfo{volume}{58}}, \bibinfo{pages}{257--265},
  \doiprefix\url{10.1080/09500340.2010.533204} (\bibinfo{year}{2011}).
\newblock \eprint{https://doi.org/10.1080/09500340.2010.533204}.

\bibitem{bondurant-93-PSK}
\bibinfo{author}{Bondurant, R.~S.}
\newblock \bibinfo{journal}{\bibinfo{title}{Near-quantum optimum receivers for
  the phase-quadrature coherent-state channel}}.
\newblock {\emph{\JournalTitle{Opt. Lett.}}} \textbf{\bibinfo{volume}{18}},
  \bibinfo{pages}{1896--1898}, \doiprefix\url{10.1364/OL.18.001896}
  (\bibinfo{year}{1993}).

\bibitem{Da_Silva2013-tk}
\bibinfo{author}{da~Silva, M.~P.}, \bibinfo{author}{Guha, S.} \&
  \bibinfo{author}{Dutton, Z.}
\newblock \bibinfo{journal}{\bibinfo{title}{Achieving minimum-error
  discrimination of an arbitrary set of laser-light pulses}}.
\newblock {\emph{\JournalTitle{Phys. Rev. A}}} \textbf{\bibinfo{volume}{87}},
  \bibinfo{pages}{052320} (\bibinfo{year}{2013}).

\bibitem{Holevo1998-mb}
\bibinfo{author}{Holevo, A.~S.}
\newblock \bibinfo{journal}{\bibinfo{title}{The capacity of the quantum channel
  with general signal states}}.
\newblock {\emph{\JournalTitle{IEEE Transactions on Information Theory}}}
  \textbf{\bibinfo{volume}{44}}, \bibinfo{pages}{269--273}
  (\bibinfo{year}{1998}).

\bibitem{Schumacher1997-cn}
\bibinfo{author}{Schumacher, B.} \& \bibinfo{author}{Westmoreland, M.~D.}
\newblock \bibinfo{journal}{\bibinfo{title}{Sending classical information via
  noisy quantum channels}}.
\newblock {\emph{\JournalTitle{Physical Review A}}}
  \textbf{\bibinfo{volume}{56}}, \bibinfo{pages}{131--138}
  (\bibinfo{year}{1997}).

\bibitem{Giovannetti2004-ki}
\bibinfo{author}{Giovannetti, V.} \emph{et~al.}
\newblock \bibinfo{journal}{\bibinfo{title}{Classical capacity of the lossy
  bosonic channel: the exact solution}}.
\newblock {\emph{\JournalTitle{Phys. Rev. Lett.}}}
  \textbf{\bibinfo{volume}{92}}, \bibinfo{pages}{027902}
  (\bibinfo{year}{2004}).

\bibitem{Giovannetti-2014-holevocapacity-thermal-loss}
\bibinfo{author}{Giovannetti, V.}, \bibinfo{author}{Garc{\'\i}a-Patr{\'o}n,
  R.}, \bibinfo{author}{Cerf, N.~J.} \& \bibinfo{author}{Holevo, A.~S.}
\newblock \bibinfo{journal}{\bibinfo{title}{Ultimate classical communication
  rates of quantum optical channels}}.
\newblock {\emph{\JournalTitle{Nat. Photonics}}} \textbf{\bibinfo{volume}{8}},
  \bibinfo{pages}{796--800} (\bibinfo{year}{2014}).

\bibitem{Chung2016-to}
\bibinfo{author}{Chung, H.~W.}, \bibinfo{author}{Guha, S.} \&
  \bibinfo{author}{Zheng, L.}
\newblock \bibinfo{journal}{\bibinfo{title}{Superadditivity of quantum channel
  coding rate with finite blocklength joint measurements}}.
\newblock {\emph{\JournalTitle{IEEE Trans. Inf. Theory}}}
  \textbf{\bibinfo{volume}{62}}, \bibinfo{pages}{5938--5959}
  (\bibinfo{year}{2016}).

\bibitem{Dutton2011-zt}
\bibinfo{author}{Dutton, Z.}, \bibinfo{author}{Guha, S.},
  \bibinfo{author}{Chen, J.} \& \bibinfo{author}{Habif, J.~L.}
\newblock \bibinfo{journal}{\bibinfo{title}{Superadditive optical
  communications with joint detection receivers and concatenated coding}}.
\newblock {\emph{\JournalTitle{Frontiers in Optics 2011/Laser Science XXVII}}}
  (\bibinfo{year}{2011}).

\bibitem{Guha2011-nr}
\bibinfo{author}{Guha, S.}, \bibinfo{author}{Dutton, Z.} \&
  \bibinfo{author}{Shapiro, J.~H.}
\newblock \bibinfo{title}{On quantum limit of optical communications:
  Concatenated codes and joint-detection receivers}.
\newblock In \emph{\bibinfo{booktitle}{2011 {IEEE} International Symposium on
  Information Theory Proceedings}}, \bibinfo{pages}{274--278}
  (\bibinfo{year}{2011}).

\bibitem{Takeoka2014-nb}
\bibinfo{author}{Takeoka, M.} \& \bibinfo{author}{Guha, S.}
\newblock \bibinfo{journal}{\bibinfo{title}{Capacity of optical communication
  in loss and noise with general quantum gaussian receivers}}.
\newblock {\emph{\JournalTitle{Phys. Rev. A}}} \textbf{\bibinfo{volume}{89}},
  \bibinfo{pages}{042309} (\bibinfo{year}{2014}).

\bibitem{Guha2011-om}
\bibinfo{author}{Guha, S.}
\newblock \bibinfo{journal}{\bibinfo{title}{Structured optical receivers to
  attain superadditive capacity and the holevo limit}}.
\newblock {\emph{\JournalTitle{Phys. Rev. Lett.}}}
  \textbf{\bibinfo{volume}{106}}, \bibinfo{pages}{240502}
  (\bibinfo{year}{2011}).

\bibitem{Wilde2013-wz}
\bibinfo{author}{Wilde, M.~M.} \& \bibinfo{author}{Guha, S.}
\newblock \bibinfo{journal}{\bibinfo{title}{Polar codes for {Classical-Quantum}
  channels}}.
\newblock {\emph{\JournalTitle{IEEE Transactions on Information Theory}}}
  \textbf{\bibinfo{volume}{59}}, \bibinfo{pages}{1175--1187}
  (\bibinfo{year}{2013}).

\bibitem{saikat-nature-photonics}
\bibinfo{author}{Chen, J.}, \bibinfo{author}{Habif, J.~L.},
  \bibinfo{author}{Dutton, Z.}, \bibinfo{author}{Lazarus, R.} \&
  \bibinfo{author}{Guha, S.}
\newblock \bibinfo{journal}{\bibinfo{title}{Optical codeword demodulation with
  error rates below the standard quantum limit using a conditional nulling
  receiver}}.
\newblock {\emph{\JournalTitle{Nature Photonics}}}
  \textbf{\bibinfo{volume}{6}}, \bibinfo{pages}{374--379},
  \doiprefix\url{10.1038/nphoton.2012.113} (\bibinfo{year}{2012}).

\bibitem{Renes_2017}
\bibinfo{author}{Renes, J.~M.}
\newblock \bibinfo{journal}{\bibinfo{title}{Belief propagation decoding of
  quantum channels by passing quantum messages}}.
\newblock {\emph{\JournalTitle{New Journal of Physics}}}
  \textbf{\bibinfo{volume}{19}}, \bibinfo{pages}{072001},
  \doiprefix\url{10.1088/1367-2630/aa7c78} (\bibinfo{year}{2017}).

\bibitem{rengaswamy2020quantummessagepassing}
\bibinfo{author}{Rengaswamy, N.}, \bibinfo{author}{Seshadreesan, K.~P.},
  \bibinfo{author}{Guha, S.} \& \bibinfo{author}{Pfister, H.~D.}
\newblock \bibinfo{journal}{\bibinfo{title}{Quantum-message-passing receiver
  for quantum-enhanced classical communications}}.
\newblock {\emph{\JournalTitle{arXiv e-prints}}}  (\bibinfo{year}{2020}).
\newblock \eprint{2003.04356}.

\bibitem{RSGP20}
\bibinfo{author}{{Rengaswamy}, N.}, \bibinfo{author}{{Seshadreesan}, K.~P.},
  \bibinfo{author}{{Guha}, S.} \& \bibinfo{author}{{Pfister}, H.~D.}
\newblock \bibinfo{title}{Quantum advantage via qubit belief propagation}.
\newblock In \emph{\bibinfo{booktitle}{2020 IEEE International Symposium on
  Information Theory (ISIT)}}, \bibinfo{pages}{1824--1829}
  (\bibinfo{year}{2020}).

\bibitem{11-qubitcomputer}
\bibinfo{author}{Wright, K.} \emph{et~al.}
\newblock \bibinfo{journal}{\bibinfo{title}{Benchmarking an 11-qubit quantum
  computer}}.
\newblock {\emph{\JournalTitle{Nature Communications}}}
  \textbf{\bibinfo{volume}{10}}, \bibinfo{pages}{5464},
  \doiprefix\url{10.1038/s41467-019-13534-2} (\bibinfo{year}{2019}).

\bibitem{chem_review}
\bibinfo{author}{Head-Marsden, K.}, \bibinfo{author}{Flick, J.},
  \bibinfo{author}{Ciccarino, C.~J.} \& \bibinfo{author}{Narang, P.}
\newblock \bibinfo{journal}{\bibinfo{title}{Quantum information and algorithms
  for correlated quantum matter}}.
\newblock {\emph{\JournalTitle{Chemical Reviews}}}
  \doiprefix\url{10.1021/acs.chemrev.0c00620} (\bibinfo{year}{2020}).

\bibitem{trappionqc}
\bibinfo{author}{Debnath, S.} \emph{et~al.}
\newblock \bibinfo{journal}{\bibinfo{title}{Demonstration of a small
  programmable quantum computer with atomic qubits}}.
\newblock {\emph{\JournalTitle{Nature}}} \textbf{\bibinfo{volume}{536}},
  \bibinfo{pages}{63--66}, \doiprefix\url{10.1038/nature18648}
  (\bibinfo{year}{2016}).

\bibitem{Monz1068}
\bibinfo{author}{Monz, T.} \emph{et~al.}
\newblock \bibinfo{journal}{\bibinfo{title}{Realization of a scalable shor
  algorithm}}.
\newblock {\emph{\JournalTitle{Science}}} \textbf{\bibinfo{volume}{351}},
  \bibinfo{pages}{1068--1070}, \doiprefix\url{10.1126/science.aad9480}
  (\bibinfo{year}{2016}).
\newblock
  \eprint{https://science.sciencemag.org/content/351/6277/1068.full.pdf}.

\bibitem{zoller-open-system-sim}
\bibinfo{author}{Barreiro, J.~T.} \emph{et~al.}
\newblock \bibinfo{journal}{\bibinfo{title}{An open-system quantum simulator
  with trapped ions}}.
\newblock {\emph{\JournalTitle{Nature}}} \textbf{\bibinfo{volume}{470}},
  \bibinfo{pages}{486--491}, \doiprefix\url{10.1038/nature09801}
  (\bibinfo{year}{2011}).

\bibitem{DJalgo}
\bibinfo{author}{Gulde, S.} \emph{et~al.}
\newblock \bibinfo{journal}{\bibinfo{title}{Implementation of the
  deutsch--jozsa algorithm on an ion-trap quantum computer}}.
\newblock {\emph{\JournalTitle{Nature}}} \textbf{\bibinfo{volume}{421}},
  \bibinfo{pages}{48--50}, \doiprefix\url{10.1038/nature01336}
  (\bibinfo{year}{2003}).

\bibitem{ULTRAFASTPhysRevLett.112.250501}
\bibinfo{author}{Muralidharan, S.}, \bibinfo{author}{Kim, J.},
  \bibinfo{author}{L\"utkenhaus, N.}, \bibinfo{author}{Lukin, M.~D.} \&
  \bibinfo{author}{Jiang, L.}
\newblock \bibinfo{journal}{\bibinfo{title}{Ultrafast and fault-tolerant
  quantum communication across long distances}}.
\newblock {\emph{\JournalTitle{Phys. Rev. Lett.}}}
  \textbf{\bibinfo{volume}{112}}, \bibinfo{pages}{250501},
  \doiprefix\url{10.1103/PhysRevLett.112.250501} (\bibinfo{year}{2014}).

\bibitem{long_distance_quantum_comm}
\bibinfo{author}{Muralidharan, S.} \emph{et~al.}
\newblock \bibinfo{journal}{\bibinfo{title}{Optimal architectures for long
  distance quantum communication}}.
\newblock {\emph{\JournalTitle{Scientific Reports}}}
  \textbf{\bibinfo{volume}{6}}, \bibinfo{pages}{20463},
  \doiprefix\url{10.1038/srep20463} (\bibinfo{year}{2016}).

\bibitem{tele-review}
\bibinfo{author}{Pirandola, S.}, \bibinfo{author}{Eisert, J.},
  \bibinfo{author}{Weedbrook, C.}, \bibinfo{author}{Furusawa, A.} \&
  \bibinfo{author}{Braunstein, S.~L.}
\newblock \bibinfo{journal}{\bibinfo{title}{Advances in quantum
  teleportation}}.
\newblock {\emph{\JournalTitle{Nature Photonics}}}
  \textbf{\bibinfo{volume}{9}}, \bibinfo{pages}{641--652},
  \doiprefix\url{10.1038/nphoton.2015.154} (\bibinfo{year}{2015}).

\bibitem{ion-photon04}
\bibinfo{author}{Blinov, B.~B.}, \bibinfo{author}{Moehring, D.~L.},
  \bibinfo{author}{Duan, L.~M.} \& \bibinfo{author}{Monroe, C.}
\newblock \bibinfo{journal}{\bibinfo{title}{Observation of entanglement between
  a single trapped atom and a single photon}}.
\newblock {\emph{\JournalTitle{Nature}}} \textbf{\bibinfo{volume}{428}},
  \bibinfo{pages}{153--157}, \doiprefix\url{10.1038/nature02377}
  (\bibinfo{year}{2004}).

\bibitem{modular_entanglement}
\bibinfo{author}{Hucul, D.} \emph{et~al.}
\newblock \bibinfo{journal}{\bibinfo{title}{Modular entanglement of atomic
  qubits using photons and phonons}}.
\newblock {\emph{\JournalTitle{Nature Physics}}} \textbf{\bibinfo{volume}{11}},
  \bibinfo{pages}{37--42}, \doiprefix\url{10.1038/nphys3150}
  (\bibinfo{year}{2015}).

\bibitem{PhysRevA.89.022317}
\bibinfo{author}{Monroe, C.} \emph{et~al.}
\newblock \bibinfo{journal}{\bibinfo{title}{Large-scale modular
  quantum-computer architecture with atomic memory and photonic
  interconnects}}.
\newblock {\emph{\JournalTitle{Phys. Rev. A}}} \textbf{\bibinfo{volume}{89}},
  \bibinfo{pages}{022317}, \doiprefix\url{10.1103/PhysRevA.89.022317}
  (\bibinfo{year}{2014}).

\bibitem{PhysRevA.52.4214}
\bibinfo{author}{Vogel, W.} \& \bibinfo{author}{Filho, R. L. d.~M.}
\newblock \bibinfo{journal}{\bibinfo{title}{Nonlinear jaynes-cummings dynamics
  of a trapped ion}}.
\newblock {\emph{\JournalTitle{Phys. Rev. A}}} \textbf{\bibinfo{volume}{52}},
  \bibinfo{pages}{4214--4217}, \doiprefix\url{10.1103/PhysRevA.52.4214}
  (\bibinfo{year}{1995}).

\bibitem{RevModPhys.75.281}
\bibinfo{author}{Leibfried, D.}, \bibinfo{author}{Blatt, R.},
  \bibinfo{author}{Monroe, C.} \& \bibinfo{author}{Wineland, D.}
\newblock \bibinfo{journal}{\bibinfo{title}{Quantum dynamics of single trapped
  ions}}.
\newblock {\emph{\JournalTitle{Rev. Mod. Phys.}}}
  \textbf{\bibinfo{volume}{75}}, \bibinfo{pages}{281--324},
  \doiprefix\url{10.1103/RevModPhys.75.281} (\bibinfo{year}{2003}).

\bibitem{2018NJPh...20d3005H}
\bibinfo{author}{{Han}, R.}, \bibinfo{author}{{Bergou}, J.~A.} \&
  \bibinfo{author}{{Leuchs}, G.}
\newblock \bibinfo{journal}{\bibinfo{title}{{Near optimal discrimination of
  binary coherent signals via atom-light interaction}}}.
\newblock {\emph{\JournalTitle{New Journal of Physics}}}
  \textbf{\bibinfo{volume}{20}}, \bibinfo{pages}{043005},
  \doiprefix\url{10.1088/1367-2630/aab2c5} (\bibinfo{year}{2018}).
\newblock \eprint{1710.04592}.

\bibitem{PhysRevA.71.023811}
\bibinfo{author}{Rodr\'{\i}guez-Lara, B.~M.}, \bibinfo{author}{Moya-Cessa, H.}
  \& \bibinfo{author}{Klimov, A.~B.}
\newblock \bibinfo{journal}{\bibinfo{title}{Combining jaynes-cummings and
  anti-jaynes-cummings dynamics in a trapped-ion system driven by a laser}}.
\newblock {\emph{\JournalTitle{Phys. Rev. A}}} \textbf{\bibinfo{volume}{71}},
  \bibinfo{pages}{023811}, \doiprefix\url{10.1103/PhysRevA.71.023811}
  (\bibinfo{year}{2005}).

\bibitem{PhysRevLett.82.1835}
\bibinfo{author}{M\o{}lmer, K.} \& \bibinfo{author}{S\o{}rensen, A.}
\newblock \bibinfo{journal}{\bibinfo{title}{Multiparticle entanglement of hot
  trapped ions}}.
\newblock {\emph{\JournalTitle{Phys. Rev. Lett.}}}
  \textbf{\bibinfo{volume}{82}}, \bibinfo{pages}{1835--1838},
  \doiprefix\url{10.1103/PhysRevLett.82.1835} (\bibinfo{year}{1999}).

\bibitem{RevModPhys.85.1083}
\bibinfo{author}{Haroche, S.}
\newblock \bibinfo{journal}{\bibinfo{title}{Nobel lecture: Controlling photons
  in a box and exploring the quantum to classical boundary}}.
\newblock {\emph{\JournalTitle{Rev. Mod. Phys.}}}
  \textbf{\bibinfo{volume}{85}}, \bibinfo{pages}{1083--1102},
  \doiprefix\url{10.1103/RevModPhys.85.1083} (\bibinfo{year}{2013}).

\bibitem{10.1117/12.2045508}
\bibinfo{author}{Boroson, D.~M.} \emph{et~al.}
\newblock \bibinfo{title}{{Overview and results of the Lunar Laser
  Communication Demonstration}}.
\newblock In \bibinfo{editor}{Hemmati, H.} \& \bibinfo{editor}{Boroson, D.~M.}
  (eds.) \emph{\bibinfo{booktitle}{Free-Space Laser Communication and
  Atmospheric Propagation XXVI}}, vol. \bibinfo{volume}{8971},
  \bibinfo{pages}{213 -- 223}, \doiprefix\url{10.1117/12.2045508}.
  \bibinfo{organization}{International Society for Optics and Photonics}
  (\bibinfo{publisher}{SPIE}, \bibinfo{year}{2014}).

\bibitem{2012arXiv1208.0391M}
\bibinfo{author}{{Monroe}, C.} \emph{et~al.}
\newblock \bibinfo{journal}{\bibinfo{title}{{Large Scale Modular Quantum
  Computer Architecture with Atomic Memory and Photonic Interconnects}}}.
\newblock {\emph{\JournalTitle{arXiv e-prints}}}
  \bibinfo{pages}{arXiv:1208.0391} (\bibinfo{year}{2012}).
\newblock \eprint{1208.0391}.

\bibitem{coupling_patent}
\bibinfo{author}{Christopher~Monroe, R.~R., Jungsang~Kim}.
\newblock \bibinfo{title}{Fault tolerant scalable modular quantum computer
  architecture with an enhanced control of multi-mode couplings between trapped
  ion qubits} (\bibinfo{year}{US9858531B1, 2018-01-02}).

\bibitem{a._kim_2018}
\bibinfo{author}{A., E. G.~A.} \& \bibinfo{author}{Kim, Y.-H.}
\newblock \emph{\bibinfo{title}{Network information theory}}
  (\bibinfo{publisher}{Cambridge University Press}, \bibinfo{year}{2018}).

\bibitem{pino2020demonstration}
\bibinfo{author}{Pino, J.~M.} \emph{et~al.}
\newblock \bibinfo{title}{Demonstration of the qccd trapped-ion quantum
  computer architecture} (\bibinfo{year}{2020}).
\newblock \eprint{2003.01293}.

\end{thebibliography}

\noindent\textbf{Author contributions}\\
S.G. suggested the idea underlying this project. P.N. and S.G. co-directed the project. C.D. and K.S. jointly worked on the theory and BPQM circuits with mid-circuit measurements. I.M., C.D. and P.N. jointly worked on the photon-ion step. C.D., I.M. and A.G. jointly performed all the experiments presented here. All authors contributed to the analysis and writing of the manuscript.\\

\noindent\textbf{Competing interests}\\
The authors declare no competing interests.\\

\section*{Methods}

\subsection*{BPQM}
The factor graph defining the 3-bit code considered in this paper is shown below:
\begin{equation}
\begin{tikzpicture}
\node[draw,circle,fill=blue!30] (x1) at (0,0) {$x_1$};
\node[draw,rectangle,fill=red!50] (c) at (0,-1) {$c$};
\node[align=center] (PC) at (+1.75,-1) {$x_1\oplus x_2\oplus x_3=0$};
\node[draw,circle,fill=blue!30] (x2) at (-0.75,-2) {$x_2$};
\node[draw,circle,fill=blue!30] (x3) at (+0.75,-2) {$x_3$};
\draw (x1) -- (c);
\draw (x2) -- (c);
\draw (x3) -- (c);
\end{tikzpicture}
\label{fig:factorgraph}
\end{equation}
which generates the set of codewords:
\begin{equation}
    \mathcal{C} = \{ 000,110,101,011 \}.
\end{equation}
The gates used for message combining at the check nodes and bit nodes of the factor graph are the Controlled-NOT gate and a unitary $U_{\ostar}$, given by
\begin{equation}
    U_{\ostar}(\theta,\theta') = \begin{pmatrix}
        a_+ & 0 & 0 & a_-\\
        a_- & 0 & 0 & -a_+\\
        0 & b_+ & b_- & 0\\
        0 & b_- & -b_+ & 0\\
        \end{pmatrix},
\end{equation}
where
\begin{equation}
a_{\pm} = \frac{1}{\sqrt{2}} \frac{ \cos{(\frac{\theta-\theta'}{2})}\pm\cos{(\frac{\theta+\theta'}{2})}}{\sqrt{1+\cos{\theta}\cos{\theta'}}} ,
\end{equation}
\begin{equation}
b_{\pm} = \frac{1}{\sqrt{2}} \frac{ \sin{(\frac{\theta+\theta'}{2})}\mp\sin{(\frac{\theta-\theta'}{2})}}{\sqrt{1-\cos{\theta}\cos{\theta'}}} ,
 \end{equation}
  \begin{equation}
\cos{\theta_0} = \frac{\cos{\theta} + \cos{\theta'}}{1+\cos{\theta}\cos{\theta'}} , \cos{\theta_1} = \frac{\cos{\theta} - \cos{\theta'}}{1-\cos{\theta}\cos{\theta'}}.
\end{equation}
In the above equations, $\theta$ captures the angle of the input qubits and can be translated to the mean photon number $N$ by the relation $e^{-2N} = \cos{\theta}$. In essence, this $U_{\ostar}$ unitary compresses the information of the two qubits into one, leaving the other in a fixed state, the $\ket{0}$ state. For more details, please refer to~\cite{rengaswamy2020quantummessagepassing, Renes_2017}.
\subsection*{Classical Limits}
%Explain classical limits in plots
When decoding the first bit alone, the ideal classical bound corresponds to performing the pulse-by-pulse detection based on the quantum optimal Helstrom measurement, followed by inference of the bit using the classical belief propagation algorithm. Since the code has a tree factor graph, classical belief propagation amounts to maximum likelihood decoding. Likewise, the practical classical bound corresponds to the same, except where the Helstrom measurement is replaced by coherent homodyne detection. The relevant pulse-by-pulse discriminating measurement average error probabilities are given by
\begin{equation}
    p_{\textrm{Hel}}=\frac{1}{2}(1-\sin\theta),\quad 
    p_{\textrm{Homodyne}}=\frac{1}{2}\operatorname{erfc}\sqrt{-\log\cos\theta}, \quad \theta\in (0,\pi/2).
\end{equation}
For the full decoder circuit, the classical bound is the average error probability associated with codeword maximum likelihood detection following either pulse-by-pulse Helstrom (ideal) or homodyne (practical) measurements.

\subsection*{Quantum Limits}
A lower bound on the quantum-enhanced classical communication capacity with the trapped-ion joint-detection receiver following photonic-to-ionic transduction, denoted as the ``achievable" Holevo capacity, is given by considering the classical-input-quantum-output (cq) channel analogue of the ``channel with random state" classical channel model, as defined in \cite{a._kim_2018}. The latter is defined as a discrete memoryless channel with state $(\mathcal{X,S},p(y|x,s),\mathcal{Y})$, with $\mathcal{X,Y,S}$ denoting the input, output and channel state alphabets, respectively (assumed to be finite), where the channel state sequence $\{S_i\}$ is an i.i.d. process with distribution $P_S(s)$, i.e., changing randomly for every use of the channel. For such a channel, there are many possible scenarios with respect to availability of the state information to the encoder and the decoder. The scenario that is relevant to us here is the one where the information about the state sequence is available only at the decoder. In this case, the capacity is given by $C=\max_p(x) I(X;Y|S)$. The achievability part follows trivially from treating $(Y^n,S^n)$ as the output of the channel $p(y,s|x)=p(s)p(y|x,s)$. The achievability holds good also when the channel output $Y$ is quantum, i.e., for a cq channel with random channel state, where the channel state is known only to the decoder. Thus, a lower bound on the achievable capacity for BPSK communications with a trapped-ion joint detection receiver goes as:
\begin{equation}
    R = P_{n=0}\times h_2\left(\frac{1 + e^{-2\eta |\alpha|^2}}{2}\right),
\end{equation}
where $P_{n=0}$ is the transduction success probability, $\eta$ is the transmissivity of the channel and $|\alpha|$ is the amplitude of the transmitted laser pulse. This value is plotted in Fig. \ref{fig:link_plot} after calculating $P_{n=0}$ for the photon-to-ion transduction mechanism discussed in the main text. We note that the converse part of the coding theorem for the cq channel remains open.

\subsection*{Link Budgets}
To describe the various link values for practical application, we calculated mean photon number values based on specs from the 2013 NASA Lunar Laser Communications Demonstration (LLCD), i.e. laser wavelength (1.6 $\mu$m), dimensions of telescopes (0.1 m on Moon/Mars and 0.4 m on Earth diameters) and laser powers (10W Uplink and 0.5 W Downlink). Additionally, we assume a modulation bandwidth of a) $\tau$ = 10 ps, i.e., 100 GHz laser source, for Moon and b) $\tau$ = 1 ns, i.e., 1 GHz pulsed laser source for Mars.

\begin{figure}
    \begin{subfigure}[b]{\textwidth}
        \caption{ } 
        \includegraphics[width=1\linewidth]{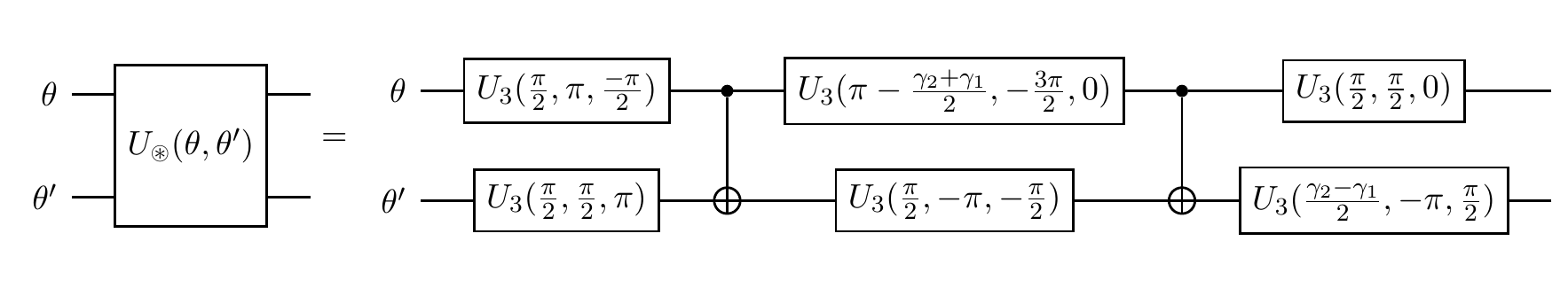}
        \label{fig:circuit_decomp}
    \end{subfigure}
    \begin{subfigure}[b]{\textwidth}
        \caption{ }
        \includegraphics[width=1\linewidth]{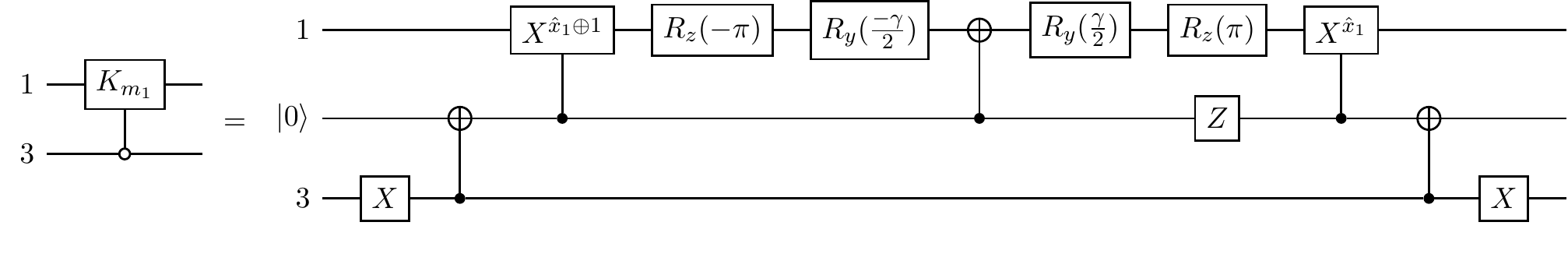}
        \label{fig:k_m}
    \end{subfigure}
    \caption{Both decompositions for BPQM full decoder components.(a) $U$ gate decomposition, where $U_3$ is the Qiskit rotation gate and $\gamma_1$,$\gamma_2$ are defined in the Eg. \ref{gammadef}. (b) $K_m$ gate decomposition, utilizing an ancilla qubit 3}
    \label{fig:full_decomps}
\end{figure}

\subsection*{Circuit Definitions and Optimizations}

For the first bit decoding in the circuit, it is equivalent to conditionally applying the two $U_{\ostar}$ gates based on an mid-circuit measurement on the third qubit following the initial CNOT gate. This avoids the trouble of decomposing $CU_{\ostar}$ into native two-qubit gates, and we can simply use $U_{\ostar}$ itself, thanks to the mid-circuit measurement capabilities of the Honeywell device. The resulting circuits have only 6 2-qubit gates. It is important to note that although each individual point's circuits were run back-to-back, all points were not collected during the same device session. Gate fidelities can drift from day-to-day on the same device, but not enough to significantly impact our results.

\noindent For the full decoder circuit, the $U_{\ostar}$ unitary was constructed by taking its components and adding a control line onto each gate, with the components shown in Fig. \ref{fig:circuit_decomp} where
 \begin{equation}
\gamma_1 = 2\sin^{-1}(a_-) , \gamma_2 = 2\sin^{-1}(b_+)
\label{gammadef}
\end{equation}

\noindent These circuit components were optimized through Qiskit's transpilation function and various pencil-and-paper optimizations, which produced a slightly different structure than the original implementation shown in \cite{rengaswamy2020quantummessagepassing}. $K_{m_1}$ is given in Fig. \ref{fig:k_m} and was applied as shown. The mid-circuit measurement on the first qubit required an active qubit reset to avoid drifting into a non-computational sub-space, and all measurements were performed in the $X$ basis. The final circuits with 2-qubit gate count of 81 were submitted to the Honeywell device via an API call to the Honeywell system in QASM form. Decomposition to native gates and qubit gate specifics were handled by Honeywell's internal software. Honeywell qubits are shuttled between various gate zones which gives effective all-to-all connectivity. See their release paper \cite{pino2020demonstration} for specifics. 

%\subsection*{Specifications of the Honeywell LT 1.0 Device}
% Was leaving this open for Honeywell people to add to if they deemed it necessary
%Description of device specs, etc

\end{document}